\pdfoutput=1

\documentclass[aps,prl,amsmath,amssymb,floatfix,twocolumn,amsmath,superscriptaddress,twocolumn,nofootinbib,tighten,letterpaper]{revtex4-2}

\usepackage[utf8]{inputenc}
\usepackage{bm}
\usepackage{siunitx}

\usepackage{enumerate}
\usepackage{amsfonts}
\usepackage{amsmath}
\usepackage{amssymb}
\usepackage{color}
\usepackage{soul}
\usepackage{float}
\usepackage{bm}
\usepackage{graphicx}
\usepackage{graphics}
\usepackage{graphicx}

\usepackage[colorlinks,allcolors=blue]{hyperref}

\let\phi=\varphi
\let\epsilon=\varepsilon

\definecolor{DarkRed}{rgb}{0.80,0,0}
\definecolor{DarkGray}{rgb}{0.7,0.7,0.7}

\newcommand{\prlsection}[1]{\textit{#1}.\kern0.05em---\kern0.05em\ignorespaces}
\usepackage[capitalise]{cleveref}

\begin{document}
\title{Fragile dislocation modes in obstructed atomic topological phases}

\author{Gabriel Malavé}
\affiliation{Facultad de Física, Pontificia Universidad Católica de Chile, Vicuña Mackenna 4860, Santiago 8331150, Chile}

\author{Jorge Schifferli}
\affiliation{Departamento de F\'isica, Universidad T\'ecnica Federico Santa Mar\'ia, Casilla 110, Valpara\'iso, Chile}

\author{Rodrigo Soto-Garrido}
\affiliation{Facultad de Física, Pontificia Universidad Católica de Chile, Vicuña Mackenna 4860, Santiago 8331150, Chile}

\author{Pedro A. Orellana}
\affiliation{Departamento de F\'isica, Universidad T\'ecnica Federico Santa Mar\'ia, Casilla 110, Valpara\'iso, Chile}

\author{Vladimir Juri\v{c}i\'c}\thanks{Corresponding author:juricic@nordita.org}
\affiliation{Departamento de F\'isica, Universidad T\'ecnica Federico Santa Mar\'ia, Casilla 110, Valpara\'iso, Chile}
\affiliation{Nordita, KTH Royal Institute of Technology and Stockholm University,
Hannes Alfvéns väg 12, SE-106 91 Stockholm, Sweden}
\begin{abstract}
We here introduce the concept of
fragile dislocation modes, which are localized only in a fraction of a topological phase while otherwise leaking into the bulk continuum.
As we demonstrate here, such dislocation modes are hosted in an obstructed atomic topological phase in the two-dimensional Su-Schrieffer-Heeger model but only in a finite region with an indirect gap at high energy.
 They are realized as chiral pairs at finite energies
 with protection stemming from a combination of the chiral (unitary particle-hole) and the point group (C$_{4v}$) symmetries, but only when the indirect gap is open.
 In this regime, we corroborate the stability of the defect modes by following their localization and also by explicitly adding a weak chemical potential disorder.
 Our findings should, therefore, be
 consequential for the experimental observation of such modes in designer topological crystals and classical metamaterials.

\end{abstract}

\maketitle

\begin{figure}[t!]
{\includegraphics[width=\linewidth]{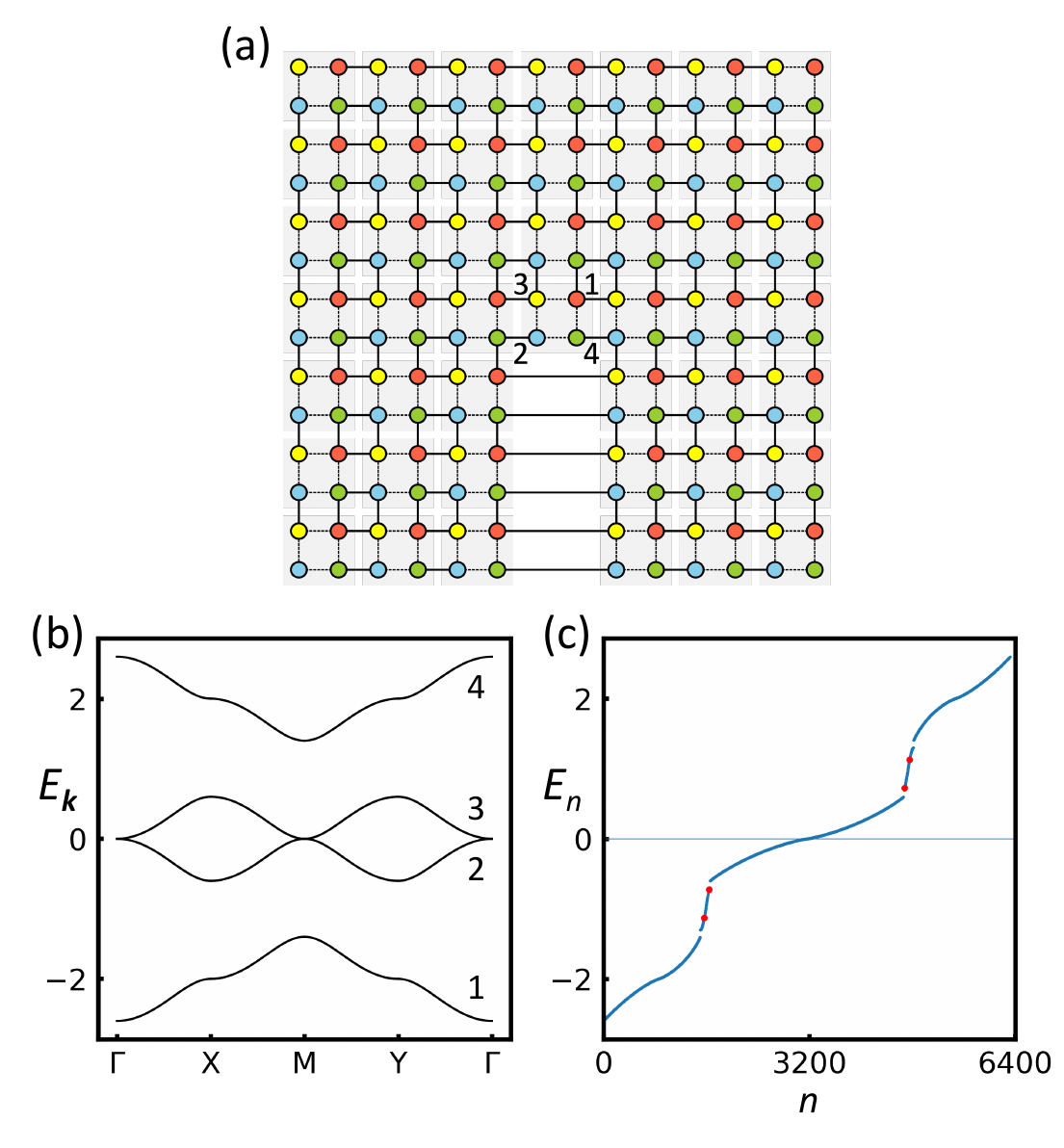}}
\caption{Two-dimensional (2D) Su-Schrieffer-Heeger (SSH) model with a single dislocation. (a) The lattice featuring a single dislocation with the Burgers vector ${\bf b}=2a{\bf e}_x$. $a$ is the distance between nearest-neighbor sites. The sites within the unit cell are labeled as $\alpha=1,...,4$, and with different colors. The dashed
 (solid) lines denote intra- (inter-)unit-cell hoppings   $t$ ($\tau$) [Eq.~\eqref{eq:Ham-SSH}]. (b) Band structure for the defect-free system along high-symmetry lines in the Brillouin zone with the corresponding spectrum displayed in Fig.~S2 of the Supplementary Materials (SM). Notice an indirect bandgap between the two positive (negative) energy bands, which closes for $|t/\tau|=0.5$, as shown in the SM, Sec.~S1. (c) The spectrum of the 2D SSH model with a single dislocation and open boundaries.  Notice two chiral pairs of dislocation modes (red marked), with energies $E=\pm 0.72$ and $E=\pm 1.13$. The corresponding LDOS  in Fig.~\ref{Fig:DisModesRealSpace} implies that these modes are localized. The system size is $40\times40$ unit cells. In (b) and (c), we set $t=0.3$ and $\tau=1$. See Fig.~S1 of the SM for additional band structure plots.}
~\label{Fig:Model}
\end{figure}



\begin{figure*}[t!]
{\includegraphics[width=1\linewidth]{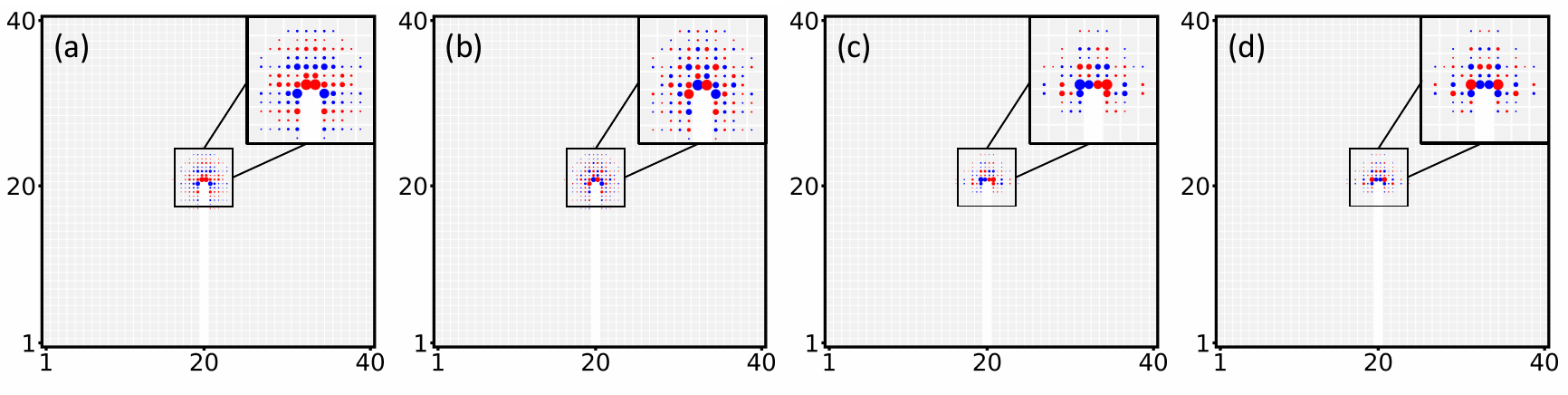}}
\caption{Two chiral pairs of the fragile dislocation bound states in the topological phase. Positive energy modes at (a) $E=0.72$ and (c) $E=1.13$, respectively,  transform under $A_1$ and $B_2$ irreducible representations of the C$_{4v}$  group while their chiral partners at (b) $E=-0.72$ and (d) $E=-1.13$, exchange the representation, i.e. transform under $B_2$ and $A_1$ representations, respectively,  therefore transforming differently under the vertical and horizontal mirror reflections, with the  $A_1$ mode being invariant, while the $B_2$ changing its sign under these operations. Notice that the modes are localized within a few unit cells in the defect core. Zoom-in of each mode is shown in the inset.  The values of the parameters are identical to Fig.~\ref{Fig:Model}. The red (blue) color denotes the positive (negative) phase of the wavefunction $\Psi_i$, while the circle's area is proportional to the amplitude $|\Psi_i|$ at the lattice site $i$. Sites with an amplitude value less than $10^{-2}$ are left empty. Consult the Supplementary Materials for additional details. }
~\label{Fig:DisModesRealSpace}
\end{figure*}



\prlsection{Introduction} The hallmark feature of topological phases are topological electronic bands characterized by the bulk topological invariants~\cite{TKNN,KM-PRL2005,Fu-Kane2006,FKM-PRL2007,moore-balents2007}, which thus cannot be localized on the real-space lattice~\cite{vanderbilt2018berry}, resulting in the hallmark bulk-boundary correspondence~\cite{hasan2010colloquium,qi2011topological}. In this case, topological lattice dislocation defects can host special topologically and symmetry-protected localized modes, providing a common mechanism for the bulk probing of topological phases covering a wide range of first-~\cite{Ran-NatPhys2009,Teo-PRB2010,Juricic-PRL2012,Asahi-PRB2012,slager-natphys2013,Slager-PRB2014,HughesYao2014,Bulmash-PRX2015,Chernodub-Zubkob2017,Kodama2019,Soto-PRR2020,Schindler2022,Panigrahi2022}, higher-order phases~\cite{Roy-Juricic-PRR2021,Schindler2022,Fulga-PRL2019,hu2022dislocation}, as well as the dynamical Floquet~\cite{Nag2021} and non-Hermitian topological systems~\cite{PanigrahiPRB2022,Schindler-PRB2021,Bhargava-PRB2021}. The  experimental observations of such dislocation modes  were reported in both quantum materials~\cite{Hamasaki-APL2017,Nayak-SciAdv2019} and classical metamaterials~\cite{Li2018,Grunberg2020,Dislocation-Acoustic2021,Ye2022,Yamada2022}. This defines the so-called bulk-dislocation correspondence, with an apparent ``duality" being at play: topological electronic wavefunctions in the bulk therefore lacking Wannier representation in the real space  yield localized modes on the dislocation defects, provided that the band inversion is at a finite momentum in the Brillouin zone (BZ)~\cite{Ran-NatPhys2009,Teo-PRB2010,Juricic-PRL2012,Asahi-PRB2012,slager-natphys2013,Slager-PRB2014,Roy-Juricic-PRR2021,Panigrahi2022,Schindler2022}. On the other hand, the absence of the localized dislocation mode does not imply topological triviality of the phase, with the paradigmatic example being the $\mathbb{Z}_2$ topological insulator featuring the band-inversion at the $\Gamma-$point~\cite{Juricic-PRL2012}.

  Obstructed atomic topological phases possess Wannier representation but at a Wyckoff position away from the lattice site~\cite{Bradlyn2017}, with one-dimensional (1D) Su-Schrieffer-Heeger (SSH) model as possibly the simplest example~\cite{SSH}. Due to this topological obstruction, they can feature charge fractionalization and nontrivial disclination responses \cite{Benalcazar-PRB2019,takahashi-2021}.   Given the existence of this class of topological phases, a natural question arises regarding the dislocation response therein since the phase admits a Wannier representation and, therefore, may obstruct the existence of the dislocation modes. In contrast, its topological nature, which can be expressed through the band representations at high-symmetry points (HSPs)~\cite{Kruthoff2017,Bradlyn2017}, is expected to yield localized dislocation modes, when the band inversion is at finite momentum. As we show here, the answer is a superposition of the two previously discussed possibilities: the dislocation modes become fragile. 

{ Two notable aspects of our results set them apart from previous studies on the lattice-dislocation modes in topological states~\cite{Ran-NatPhys2009,Teo-PRB2010,Juricic-PRL2012,Asahi-PRB2012,slager-natphys2013,Slager-PRB2014,Roy-Juricic-PRR2021,Panigrahi2022,Schindler2022}. First, we demonstrate the existence of  \emph{fragile dislocation modes}, which are localized and stable but only in a particular region within the topological phase characterized by an indirect bandgap, and exemplify it in the case of two-dimensional (2D) SSH model. Second, we show that their stability is due to  the subtle interplay between the lattice symmetry, electronic topology and the form of spectral (indirect) gap in the obstructed topological state. As such, fragile dislocation modes can serve as bulk probes of the obstructed  topological phase but only in its part  featuring the indirect bandgap. In the following, we provide a detailed summary of our key results.


\prlsection{Key results} 
By performing numerical analysis of a 2D SSH model featuring a paradigmatic obstructed topological phase~\cite{Liu-PRL2017,Obana-PRB2019,Benalcazar-CerjanPRB2020},  we find that a single dislocation can bind several pairs of modes at finite energies throughout this phase  (Fig.~\ref{Fig:Model}), which are localized (Fig.~\ref{Fig:DisModesRealSpace}) due to the dislocation-flux correspondence~\cite{Roy-Juricic-PRR2021,Schindler2022}, and symmetry protected against hybridization with each other. Crucially, numerical analysis of the evolution of the modes' localization throughout the topological phase demonstrates their stability (Fig.~\ref{Fig:DisModesEvolution}), but only in its finite region, with the modes dissolving into the bulk when a high-energy indirect gap closes, and the system becomes gapless, while remaining topological,  see also Fig.~\ref{Fig:EnergyEvolution}. Furthermore,  when localized, the dislocation modes are robust with respect to a weak chemical-potential disorder (Fig.~\ref{Fig:DisModesDisorder}). Finally, possible experimental relevance of these fragile dislocation modes in metamaterials and designer materials platforms is outlined, particularly in a silicon-based designer lattice~\cite{Geng2022} and a silicon dielectric metasurface platform~\cite{Doiron2022}.

\prlsection{2D SSH model.} We start with the Bloch Hamiltonian of the 2D SSH model on a square lattice, $\hat{H}=\sum_{\mathbf{k}}{{\Psi}}^\dagger_{\mathbf{k}}\mathbb{H}_{\mathbf{k}}{{\Psi}}_{\mathbf{k}}$, where ~\cite{Liu-PRL2017} 
\begin{equation}
\label{eq:Ham-SSH}
\mathbb{H}_{\bf k}=\sum_{i,j=0}^3 A_{ij}({\bf k})\sigma_i\otimes\sigma_j,
\end{equation}
with the nonvanishing form factors given by 
\begin{align}
&A_{10}=t+\tau\cos(2k_xa),\,\, A_{11}=t+\tau\cos(2k_ya),\nonumber\\  
&A_{23}=-\tau\sin(2k_xa),\,\,  A_{12}=-\tau\sin(2k_ya).
\end{align}
Here,   $\Psi_{\bf k}=(c_{1,{\bf k}},c_{2,{\bf k}},c_{3,{\bf k}},c_{4,{\bf k}})^\top$  represents the annihilation operator for the states at four sites in the unit cell [Fig.~\ref{Fig:Model}(a)], 
with ${\bf k}$ as the momentum, and $a$ is the nearest-neighbor distance. Parameter $t$ ($\tau$) represents the intra- (inter-)unit-cell hopping amplitude. We fix $\tau=1$ hereafter,   $\sigma_0$ is the $2\times2$ identity matrix, and $\sigma_i$ ($i=1,2,3$) are the usual Pauli matrices.  See Sec.~S1 of the Supplemental Materials (SM)~\cite{SM}.  Importantly, as shown in  in Fig.~\ref{Fig:Model}(b),  high-energy bands are separated by an indirect bandgap from the low-energy ones for $|t/\tau|<0.5$. Furthermore, this Hamiltonian features $C_{4v}$ point-group, and the chiral (unitary particle-hole) symmetry, $\{\Pi,\mathbb{H}_{\bf k}\}=0$,  generated by $\Pi=\sigma_3\otimes\sigma_0$. This symmetry implies that the eigenstates of the Hamiltonian~\eqref{eq:Ham-SSH} forming a chiral pair  $(\Psi_{\bf
k},\Pi\Psi_{{\bf k}})$ are with the same absolute value but the opposite signs of energy, $\mathbb{H}_{\bf k}\Pi^n\Psi_{\bf k}=(-1)^nE_{\bf k}\Pi^n\Psi_{\bf k}$ ($n=0,1$), and the chiral operator exchanges the irreducible representations (irreps) of the $C_{4v}$ group, $A_1$ into $B_2$, and vice versa.

\begin{table}
\begin{tabular}{|c|c|c|c|c|c|c|c|}
\hline
Phase & Bands & \multicolumn{2}{|c|}{$C_{4v}$  {\rm HSPs}} & \multicolumn{3}{|c|}{$C_{2v}$ {\rm HSPs}}  \\  \hline
 &  & $\Gamma$ & $M$ & $\Gamma$ & $X$ & $Y$  \\ \hline
\textbf{Topological} & 1 & $B_2$ & $A_1$ & $a_2$ & $b_1$ & $a_2$  \\ 
 $|t/\tau|<1$     & 2,3 & $E$ & $E$ & $b_1+b_2$ & $a_1+a_2$ & $a_1+a_2$  \\
      & 4  & $A_1$ & $B_2$ & $a_1$ & $b_2$ & $b_1$  \\ \hline \hline 
\textbf{Trivial} & 1 & $B_2$ & $B_2$ & $a_2$ & $a_2$ & $a_2$ \\
 $|t/\tau|>1$     & 2,3 & $E$ & $E$ & $b_1+b_2$ & $b_1+b_2$ & $b_1+b_2$  \\
      & 4  & $A_1$ & $A_1$ & $a_1$ & $a_1$ & $a_1$   \\ \hline
      \end{tabular}
\caption{Irreducible representations of the electronic bands in the topological and trivial phase of the  2D SSH model [Eq.~\eqref{eq:Ham-SSH}] at the high-symmetry points (HSPs) in the Brillouin zone. The bands are labelled as shown in Fig.~\ref{Fig:Model}(b). The parameters $t$ and $\tau$ represent, respectively, intra-unit-cell and inter-unit-cell hoppings, see also Eq.~\eqref{eq:Ham-SSH}. Notice that only the  bands $2$ and $3$ touching at zero energy at the $\Gamma$ and $M$ points transform under the two-dimensional $E$ representation of the $C_{4v}$ group at these HSPs. All other band representations are one-dimensional.}
\label{Tab:Irreps-HSPs}
\end{table}

Topological phases in the 2D SSH model [Eq.~\eqref{eq:Ham-SSH}] are  distinguished by the relative difference of the  irreps of high-energy electronic bands at the HSPs $M$ and $X(Y)$ with respect to the $\Gamma$ point~\cite{Obana-PRB2019}. In a topologically trivial phase, with the Wyckoff position centered at the site of the unit cell,  the irreps at the HSPs are equal, as explicitly shown in Table~\ref{Tab:Irreps-HSPs}.   In the topological phase, on the other hand, the band irreducible representations of $C_{4v}$ ($C_{2v}$) point group at the $M$ ($X$ and $Y$) point(s) differ relatively to the $\Gamma$ point, moving the Wyckoff position to the center of the unit cell (away from any atomic position) 
\cite{remark},  see also Table~\ref{Tab:Irreps-HSPs}.  Importantly, only  the two closest-to-zero energy bulk bands touching at the $M$ and $\Gamma$ points transform under the 2D $E$ representation of  $C_{4v}$ group at these HSPs. This then directly pertains to the irrep content of the the corner modes, with one pair transforming  under the $E$ irrep while the other two such modes belong to $A_1$ and $B_2$ irreps~\cite{Benalcazar-CerjanPRB2020}. 
The gap between the bulk and the edge states closes at $|t|=1$, separating topological and trivial (featureless metallic) phase. 


\prlsection{Fragile dislocation modes in the 2D SSH model} We now include a dislocation defect in the lattice [Fig.~\ref{Fig:Model}(a)], with the Burgers vector ${\bf b}=2a{\bf e}_x$, equal to a primitive lattice vector. Since the lowest-energy bands touch at the $M$  point, the dislocation  acts as an effective $\pi$ flux,   $\Phi_{\rm dis}={{\bf K}_M}\cdot{\bf b}=\pi$~\cite{Ran-NatPhys2009,Teo-PRB2010,Juricic-PRL2012}. Furthermore, the Burgers vector is orthogonal to the gapped edges, and the dislocation therefore binds finite-energy modes~\cite{Roy-Juricic-PRR2021,Schindler2022}, which, as we show here, remain stable until the indirect bandgap closes, and  dissolve into the bulk continuum.

To this end, we numerically diagonalize  the 2D SSH model in Eq.~\eqref{eq:Ham-SSH}, on a real space lattice with a single dislocation and open boundary conditions, see Fig.~\ref{Fig:Model}(a),  using the Kwant code~\cite{kwant}. In the spectrum, displayed in Fig.~\ref{Fig:Model}(c), we show two chiral pairs of modes at finite energies, representing two localized dislocation modes. The corresponding local density of states (LDOS) is displayed in  Fig.~\ref{Fig:DisModesRealSpace}, and shows that the modes within the chiral pair transform under $A_1$ and $B_2$ representations of the $C_{4v}$ group (see also Sec.~S2 and Fig.~S3 in the SM~\cite{SM}). Furthermore, the defect modes with the same sign of energy transform under different 1D irreps. 
The modes are buried within the edge continuum, separated by an indirect gap from the bulk  [Fig.~\ref{Fig:Model}(c) and Fig.~S4 in the SM~\cite{SM}], and as such protected from the mixing with  bulk modes. The hybridization with edge modes is avoided due to the geometry (bulk-edge separation). However, as the indirect gap starts to close, the two bulk bands approach the energy of the dislocation modes, and  eventually hybridize at the gap closing,  $t=0.5$, since the dislocation and (some of) the bulk modes become degenerate and transform under the same irreps, $A_1$ and $B_2$.

\prlsection{Stability of the dislocation modes} To establish stability of the modes, we follow the evolution of the localization of a representative mode throughout the topological phase, see Fig.~\ref{Fig:DisModesEvolution}.  Notice first that when the indirect gap is open, spectral weight of the mode is peaked around the defect center [Fig.~\ref{Fig:DisModesEvolution}(a)-(d)], see also Fig.~\ref{Fig:DisModesRealSpace}(a) for the mode's real-space profile  when  $t=0.3$. Once  the indirect-bulk-gap closing ($t=0.5$) is approached, the maximum spectral weight decreases significantly, as can be seen in Fig.~\ref{Fig:DisModesEvolution}(e).  Upon entering the gapless regime, the dislocation modes dissolve into the bulk,  Fig.~\ref{Fig:DisModesEvolution}(f), due to the hybridization. See also Fig.~S5 in the SM~\cite{SM} for the evolution of the defect mode's LDOS.

The indirect gap and the energy splitting between the two dislocation modes exhibit an analogous behavior, and the dislocation defect may serve as a bulk probe of the bulk-boundary correspondence in the topological phase through these fragile topological modes but only up to the closing of indirect bandgap. In Fig.~\ref{Fig:EnergyEvolution}, we observe that the gap between the two dislocation modes evolves almost linearly with the ratio $t/\tau$, analogously  to the indirect bandgap, 
$\delta E_{{\rm ind}}/\tau=2-4t/\tau$, as shown in Sec.~S1 of the SM~\cite{SM}. However, the dislocation modes are stable only up to the closing of the indirect gap, taking place at $|t/\tau|=0.5$, while the corner modes remain localized until the topological transition takes place at $|t/\tau|=1$, as shown in Fig.~S6 in the SM~\cite{SM}.



\begin{figure*}[t!]
\includegraphics[width=1\linewidth]{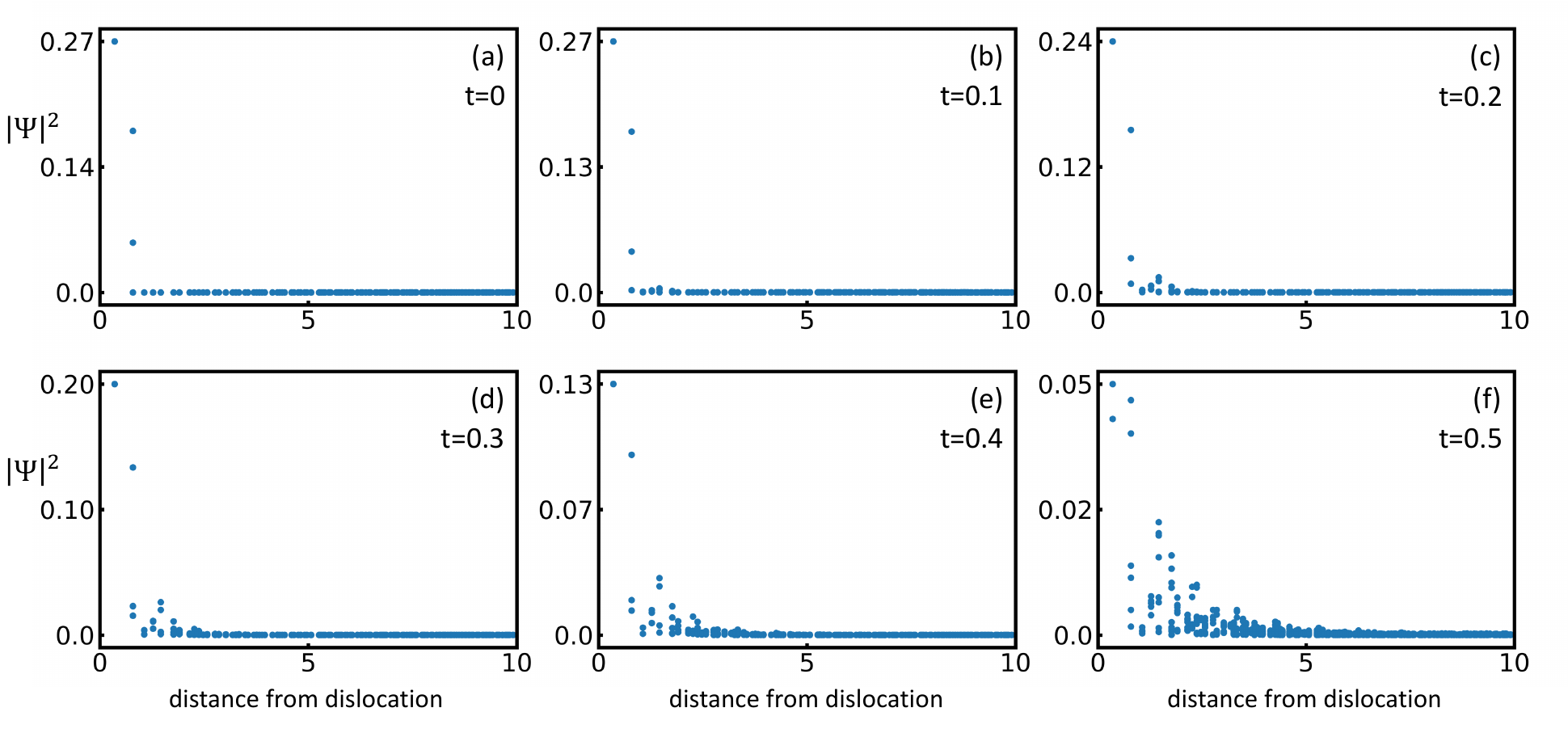}\hspace{.1cm}
\caption{Localization of the closest-to-zero positive-energy dislocation mode in the topological phase. The local density of states (LDOS) of the mode is shown for the following values of the intra-unit-cell hopping parameter $t$ [Eq.~\eqref{eq:Ham-SSH}]: (a) $t=0$; (b) $t=0.1$; (c) $t=0.2$; (d) $t=0.3$; (e) $t=0.4$; (f) $t=0.5$. We fix $\tau=1$, the system size is $40\times40$ unit cells, and the LDOS is shown only up to $10$ lattice sites from the dislocation center (LDOS farther away is negligible). Distance is given in units of the nearest-neighbor spacing, $a$. The LDOS profile of the mode for $t=0.3$ is shown in Fig.~\ref{Fig:DisModesRealSpace}(a). See Sec.~S4 of the Supplementary Materials for analogous plots of the corner modes. }
~\label{Fig:DisModesEvolution}
\end{figure*}



\begin{figure}[t!]
{\includegraphics[width=.8\linewidth]{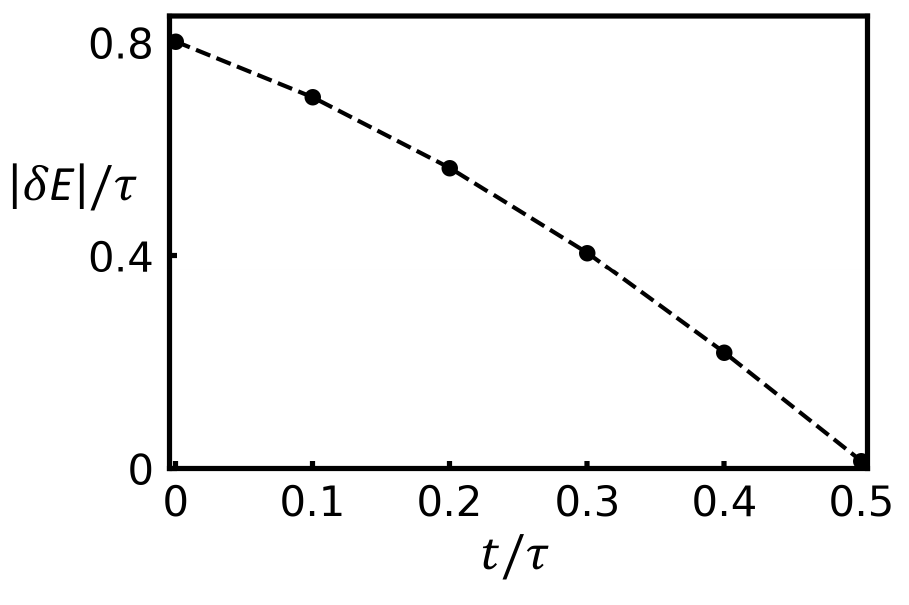}}

\caption{Evolution of the energy difference between the two fragile dislocation modes with positive energy that remain localized up to the indirect bandgap closing in the topological phase. The system size is $40\times40$ unit cells, and open boundary conditions are employed.}
~\label{Fig:EnergyEvolution}
\end{figure}


We furthermore analyze the stability of the modes with respect to the chemical-potential disorder on average preserving both chiral and $C_{4v}$ symmetries. 
We clearly observe that the modes remain localized for a weak chemical-potential disorder,  Fig.~\ref{Fig:DisModesDisorder}(a), while for the 
strong disorder, they delocalize into the bulk, see Fig.~\ref{Fig:DisModesDisorder}(b).


\emph{Mechanism of the modes' emergence and dissolution}. To explain the emergence of dislocation modes, we invoke the chiral and $C_{4v}$ symmetries of the 2D SSH model [Eq.~\eqref{eq:Ham-SSH}], which, in fact, \emph{guarantee} their existence, together with dislocation-flux correspondence~\cite{Roy-Juricic-PRR2021}. To this end, we first recall that the dislocation modes form by the Volterra process connecting the two edges across the removed line of unit cells ending at the dislocation center [Fig.~\ref{Fig:Model}(a)]. The low-energy modes localized at the two sides of the ``trench", formed before the dislocation is constructed, hybridize to yield well localized dislocation modes. The 2D SSH topological metal has two zero-energy (finite-energy) modes at the lower (upper) corners of the trench,  shown in Fig.~S3 of the SM~\cite{SM}, for a representative value $t=0.3$.
Two pairs of the top trench modes transform under even ($A$) and odd ($B$) representations of the $C_2$ group, which become one-dimensional $A_1$ and $B_2$ irreps of the $C_{4v}$  group after reconnecting the sides of the trench, and thereby restore $C_{4v}$ symmetry, see also Sec.~S2 of the SM~\cite{SM}. This irrep content is fixed by the invariance of the modes under the two diagonal reflections in the unit cell. After the edges are reconnected, the dislocation modes are at finite energy due to the hybridization and the $\pi-$flux localization, with the chiral symmetry forcing them to organize in chiral pairs. Furthermore, the chiral pairs, in general, are with absolute values of energy $\epsilon_1 > \epsilon_0$, where $\epsilon_1$ ($\epsilon_0$) corresponds to the finite (zero) energy modes \emph{before} the reconnection. If positive energy dislocation modes transform under  $B_2$ and $A_1$ irreps, their chiral partners at negative energies swap the irreps (transform under $A_1$ and $B_2$). Therefore, the modes are pairwise protected, independently of the energy splitting between the modes.  Now, the crucial part is that the high-energy bulk bands near the  $M$ point also transform under the same $A_1$ and $B_2$ representations of the $C_{4v}$ group, and therefore dislocation modes are protected from hybridization as long as the indirect gap between the positive energy bands is finite. Once this gap closes, the modes  dissolve in the bulk since the spectral gap protection is no longer operative, as explicitly shown in Fig.~S5 in SM~\cite{SM}.  Finally, even though the dislocation modes transform under the same irreps as the edge modes, the mixing between them is suppressed by the geometry (bulk-edge separation).

\prlsection{Experimental feasibility} The prime candidate for the realization of the proposed fragile dislocation modes is the Si lattice, recently shown to host the 2D SSH model in a rectangular geometry~\cite{Geng2022}. Such tunable lattice geometry, in particular, should make it possible to readily implement a dislocation defect and probe the defect modes using scanning tunneling spectroscopy.  Furthermore, engineered metasurfaces, such as  a recently reported  silicon-based system~\cite{Doiron2022},  provide another viable path to their realization via hopping control.
Photonic ~\cite{Li2018,BIC-Rechtsman-PRL2020} and acoustic crystals~\cite{ZhengPRApp2019},  mechanical resonator networks~\cite{Grunberg2020}, and topolectric circuits~\cite{Imhof2018,Bao-PRB2019,dongjuricicroy:2021PRR,Yang2022}, represent additional potential platforms for their realization. 

\prlsection{Discussion \& outlook}
In this paper, we have demonstrated that dislocation defects in an obstructed atomic topological phase can host fragile dislocation modes: the defect modes which are stable, but only in a finite fraction of the topological phase. Furthermore, our findings shed light on the importance of spectrum structure within the same topological phase for defect response.

Our findings are expected to corroborate future studies of the stability of obstructed atomic topological phases against disorder and interactions. Their generalization to higher-dimensional SSH models~\cite{liu2023analytic},  where crystalline, particle-hole and time-reversal symmetries may provide further protection to the fragile dislocation modes, is an interesting prospect for future investigation, relevant for quantum materials, e.g., bismuth~\cite{Kim-PRB2023}.

In closing, our work leaves an open question regarding the possible existence of the symmetry-protected dislocation bound-states in the continuum (BICs)~\cite{BIC-Review2016}. This issue is especially pertinent given that BICs are relevant in a variety of topological condensed-matter systems~\cite{Yang2013,Murakami-PRB2019,Grez-PRA2022,Pu-PRL2023,Li-Li-PRL2023,Yan:23}. 
Further research into the bulk-defect correspondence in obstructed atomic topological phases, driven by our findings, may uncover novel observable consequences, e.g. experimental probes, of  topological classifications based on the symmetry content of electronic bands~\cite{Kruthoff2017,Bradlyn2017}.

\prlsection{Acknowledgments} The authors are thankful to Bitan Roy and Wladimir Benalcazar for useful discussions. This work was supported by ANID/ACT210100 (G.M., R.S.-G. and V.J.), the Swedish Research Council Grant No. VR 2019-04735 (V.J.), Fondecyt (Chile) Grants No. 1200399 (R.S.-G.), No. 1201876 (P.A.O. and J. S.)  and No. 1230933 (V.J. and P. A. O.).

\begin{figure}[t!]
{\includegraphics[width=\linewidth]{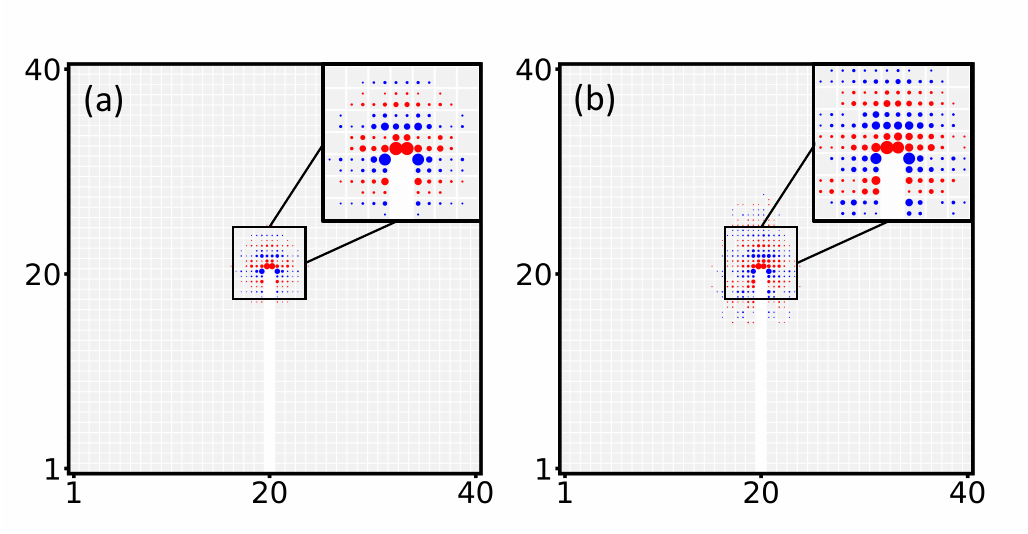}}
\caption{Effect  of the random chemical potential  on the local density of states corresponding to the dislocation modes in Fig.~\ref{Fig:DisModesRealSpace}(a).  We implement the chemical-potential $\mu$ such that at a lattice point $i$, $\mu_i = \mu(1+\delta_i) $
where the random variable $\delta_i \in [-x,x]$ is uniformly distributed, and  $x$ as the variation. We take a fixed variation, $x=0.1$, and the values of chemical potential (a) $\mu=0.3$ and (b) $\mu=3.0$. The color code is the same as in Fig. \ref{Fig:DisModesRealSpace}, sites with an amplitude $<10^{-2}$ are left empty. Zoom-in of each mode is shown in the inset.}
\label{Fig:DisModesDisorder}
\end{figure}


%

\end{document}